# Constraints on the Microphysics of Pluto's Photochemical Haze from *New Horizons* Observations

Peter Gao[1,]*, Siteng Fan[1], Michael L. Wong[1], Mao-Chang Liang[2], Run-Lie Shia[1], Joshua A. Kammer[3], Yuk L. Yung[1], Michael E. Summers[4], G. Randall Gladstone[5,6], Leslie A. Young[3], Catherine B. Olkin[3], Kimberly Ennico[7], Harold A. Weaver[8], S. Alan Stern[3], and the New Horizons Science Team

[1]*Division of Geological and Planetary Sciences, California Institute of Technology, Pasadena, CA, USA, 91125*

[2]*Research Center for Environmental Changes, Academia Sinica, Taipei 115, Taiwan*

[3]*Southwest Research Institute, Boulder, CO 80302, USA*

[4]*George Mason University, Fairfax, VA 22030, USA*

[5]*Southwest Research Institute, San Antonio, TX 78238, USA*

[6]*University of Texas at San Antonio, San Antonio, TX 78249, USA*

[7]*National Aeronautics and Space Administration, Ames Research Center, Space Science Division, Moffett Field, CA 94035, USA*

[8]*The Johns Hopkins University Applied Physics Laboratory, Laurel, MD 20723, USA*

*Corresponding author at: *Division of Geological and Planetary Sciences, California Institute of Technology, Pasadena, CA, USA, 91125*

*Email address:* pgao@caltech.edu

*Phone number:* 626-298-9098

**Abstract**

The *New Horizons* flyby of Pluto confirmed the existence of hazes in its atmosphere. Observations of a large high- to low- phase brightness ratio, combined with the blue color of the haze (indicative of Rayleigh scattering), suggest that the haze particles are fractal aggregates, perhaps analogous to the photochemical hazes on Titan. Therefore, studying the Pluto hazes can shed light on the similarities and differences between the Pluto and Titan atmospheres. We model the haze distribution using the Community Aerosol and Radiation Model for Atmospheres assuming that the distribution is shaped by downward transport and coagulation of particles originating from photochemistry. Hazes composed of both purely spherical and purely fractal aggregate particles are considered. General agreement between model results and solar occultation observations is obtained with aggregate particles when the downward mass flux of photochemical products is equal to the column-integrated methane destruction rate ~$1.2 \times 10^{-14}$ g $cm^{-2}$ $s^{-1}$, while for spherical particles the mass flux must be 2-3 times greater. This flux is nearly identical to the haze production flux of Titan previously obtained by comparing microphysical model results to Cassini observations. The aggregate particle radius is sensitive to particle charging effects, and a particle charge to radius ratio of 30 $e^-/\mu m$ is necessary to produce ~0.1-0.2 μm aggregates near Pluto's surface, in accordance with forward scattering measurements. Such a particle charge to radius ratio is 2-4 times higher than those previously obtained for Titan. Hazes composed of spheres with the same particle charge to radius ratio have particles that are 4 times smaller at Pluto's surface. These results further suggest that the haze particles are fractal aggregates. We also consider the effect of condensation of HCN, $C_2H_2$, $C_2H_4$, and $C_2H_6$ on the haze particles, which may play an important role in shaping their altitude and size distributions.

## 1. INTRODUCTION

Photochemical hazes naturally arise in reducing atmospheres due to the destruction of methane and higher hydrocarbons by solar UV photons and high-energy ions and neutrals, followed by polymerization of the resulting radical species. Such conditions have been hypothesized to exist in the atmosphere of Pluto from detections of spectral features belonging to methane ice on its surface (e.g. Cruikshank & Brown, 1986; Owen et al., 1993), which suggests the possible existence of atmospheric hazes. However, previous attempts at characterizing the lower atmosphere of Pluto by exploiting the "knee" in stellar occultation light curves (where the slope of ingress/egress steepens closer to Pluto's surface) have been inconclusive, as both a strong thermal inversion and a haze layer can produce such an observation (e.g. Elliot et al., 1989; Stansberry et al., 1994; Young et al., 2008; Rannou & Durry, 2009; Person et al., 2013; Olkin et al., 2014; Dias-Oliveira et al., 2015; Gulbis et al., 2015).

The July 14$^{th}$, 2015 flyby of Pluto by the *New Horizons* probe settled the debate by confirming the existence of optically thin haze layers pervading the lower few hundred kilometers of Pluto's atmosphere, as seen in both forward scattering observations and solar/stellar occultations (Stern et al., 2015; Gladstone et al., 2016). The haze appeared blue, which is likely due to Rayleigh scattering by small particles. However, it also featured a large high- to low-phase brightness ratio, indicative of large particles. These two observations combined point to the haze particles being fractal aggregates, drawing strong parallels between the Pluto haze and the hazes of Titan (Elliot et al., 2003). Thus, studying the Pluto haze can greatly inform comparisons of these two apparently disparate worlds. In addition, the haze particles can act as nucleation sites for condensable species in Pluto's atmosphere, thereby directly affecting atmospheric chemistry. In this work, we compare the observed haze extinction profiles with those calculated by microphysical models, thereby offering insights into the major processes controlling the haze distribution.

We describe our microphysical model in Section 2. In Section 3, we present our model results, compare them to data, and discuss the impact of condensation on the haze distribution.

## 2. MODEL

We model the Pluto haze as a direct analog of the Titan haze, as the atmospheric chemistry of the two worlds is similar (Wong et al., 2015), and uncertainties on the processes governing the formation and evolution of Titan's haze is much smaller than that of Pluto's haze owing to 10 years of Cassini observations. Possible differences between the hazes on these two worlds are discussed in Section 3.1.

Titan's haze is composed of fractal aggregates – fluffy, porous particles consisting of smaller subunits, or "monomers" (West et al., 1991). The number of monomers, $N_m$, in an aggregate particle is defined as

$$N_m = \left(\frac{R_f}{r_m}\right)^{D_f} \tag{1}$$

where $R_f$ is the effective particle radius of the aggregate, $r_m$ is the monomer radius, and $D_f$ is the fractal dimension of the aggregate, typically between 1.75 and 2.5 for Titan haze particles (Cabane et al., 1993). For simplicity we use $D_f = 2$ in our modeling. Eq. (1) can be used to find the mass $M$ of an aggregate, which for $D_f = 2$ is

$$M = \frac{4}{3}\pi\rho_p r_m R_f^2 \tag{2}$$

where $\rho_p$ ~ 1 g cm$^{-3}$ is the monomer mass density (Lavvas et al., 2010). Gladstone et al. (2016) estimated $r_m$ to be ~10 nm for Pluto's haze particles using Rayleigh scattering, and thus we adopt this value for our nominal model. We also consider 5 nm monomers, under the hypothesis that monomers can grow by condensation of simple hydrocarbons and nitriles (Barth, 2014; Wong et al., this issue).

We use the same model atmosphere as for the photochemical calculations of Wong et al., (this issue), though we focus only on altitudes below 500 km to avoid spending computational time on altitudes where data quality is low. Figure 1 shows the temperature profile of the model atmosphere, updated from Zhu et al. (2014).

We calculate the equilibrium haze particle size and altitude distribution using the 1-dimensional Community Aerosol and Radiation Model for Atmospheres (CARMA; Turco et al., 1979; Toon et al., 1988; Jacobson & Turco, 1994; Bardeen et al., 2008; Wolf & Toon, 2010). CARMA is a Eulerian forward model that solves the discretized continuity equation for aerosol particles subject to vertical transport and production and loss due to particle nucleation, condensation, evaporation, and coagulation, and has been used to simulate the Titan hazes (e.g. Toon et al., 1992; Larson et al., 2014; 2015). Coagulation and transport of Titan's haze particles dominate over nucleation and condensation/evaporation above the troposphere (Lavvas et al., 2010; Larson et al., 2014), and thus the change with time of $n_p(z)$, the number density of particles in the pth mass bin at altitude $z$ can be simplified to

$$\frac{\partial n_p}{\partial t} = -\frac{\partial}{\partial z}\left(w_{sed} n_p\right) + \frac{1}{2}\sum_{i=1}^{i=p-1} K_{i,p-i}\, n_i n_{p-i} - n_p \sum_{i=1}^{i=N} K_{i,p} n_i \quad (3)$$

where $w_{sed}$ is the sedimentation velocity and $K_{i,j}$ is the Brownian coagulation kernel between particles in mass bins $i$ and $j$. The mass bins are set such that each bin corresponds to particle masses twice that of the previous bin. We use 30 bins in our model, starting from a minimum radius of $R_f = 20$ nm, which was chosen such that the smallest particles are still fractal in nature. A fixed downward mass flux of these minimum-$R_f$ aggregate particles (with $r_m$ = 5 or 10 nm) is imposed at the top boundary of the model atmosphere, while at the lower boundary the aggregates are assumed to sediment out onto the surface.

The first term on the right hand side of Eq. (3) represents vertical transport due to sedimentation of the aerosol particles; the second term represents the increase in $n_p$ due to the coagulation of smaller particles with total mass equal to that of particles in the pth mass bin; and the third term represents the decrease in $n_p$ due to coagulation of particles in the pth mass bin with other particles to generate more massive particles. All variables in Eq. (3) are functions of $z$.

For non-porous spherical particles of radius $r_p$, $w_{sed}$ is given by

$$w_{sed} = \frac{2}{9}\frac{\rho_p g r_p^2 f_s}{\eta} \quad (4)$$

where $g$ is the gravitational acceleration of Pluto (61.7 cm $s^{-1}$ at the surface; Stern et al., 2015), $\eta$ is the dynamic viscosity, in units of Poise, calculated using Sutherland's formula as a function of the temperature $T$ and the appropriate constants for pure $N_2$ gas (White, 1991)

$$\eta = 1.781 \times 10^{-4} \frac{411.55}{T+111}\left(\frac{T}{300.55}\right)^{1.5} \quad (5)$$

and $f_s$ is the Cunningham slip correction factor, given by

$$f_s = 1 + 1.246Kn + 0.42Kne^{-0.87/Kn} \quad (6)$$

(Pruppacher & Klett, 1978) where $Kn$ is the Knudsen number defined as the ratio of the atmospheric mean free path $l$ to $r_p$, where $l$ can be written as

$$l = \frac{2\eta}{\rho}\sqrt{\frac{\pi\mu}{8RT}} \quad (7)$$

where $\rho$ is the atmospheric mass density, $\mu$ is the mean molecular weight of the atmosphere, and $R$ is the universal gas constant.

$f_s$ ~ 1 in the continuum regime ($Kn$ << 1); however, at Pluto's surface, where $T$ ~ 40 K and the atmospheric pressure ~ 10 $\mu$bar (Stern et al. 2015), $Kn$ ~ 3000 for $r_p$ ~ 0.1 $\mu$m, and thus $f_s$ is large and linear with $Kn$ (kinetics regime). Taking this into account, Eq. (4) becomes

$$w_{sed} = A\frac{\rho_p g r_p}{\rho}\sqrt{\frac{\pi\mu}{2RT}} \tag{8}$$

where A is a constant that is ~0.5. Eq. (8) is the dominant form of the sedimentation velocity throughout Pluto's atmosphere for $r_p$ < 300 $\mu$m.

Transport by diffusion is not included in the model as it is minor in comparison to sedimentation. The eddy diffusion coefficient $K_{zz}$ has been constrained to be ~1000 cm$^2$ s$^{-1}$ by Wong et al. (this issue) for all altitudes, while the Brownian diffusion coefficient $D$ can be defined as

$$D = \frac{B}{r_p^2 n_a}\sqrt{\frac{RT}{2\pi\mu}} \tag{9}$$

in the kinetics limit, where $n_a$ is the number density of air molecules and $B$ is a numerical constant that is ~0.25 (Pruppacher & Klett, 1978). To compare transport by sedimentation, eddy diffusion, and Brownian diffusion, we define a time scale $\tau$ = the time needed to traverse $d$ = 1 km in the atmosphere. For sedimentation $\tau$ ~ $d/w_{sed}$; for eddy diffusion $\tau$ ~ $d^2/K_{zz}$ = $10^7$ s; and for Brownian diffusion $\tau$ ~ $d^2/D$. Figure 2 shows $\tau$ of these three transport processes for $R_f$ = 0.1 $\mu$m, $r_m$ = 10 nm, and $D_f$ = 2. It is clear that sedimentation occurs the fastest, and therefore dominates the transport in Pluto's atmosphere. This outcome is not changed when $R_f$ and $r_m$ are altered within the ranges considered in this work.

The decrease of $g$ with altitude is taken into account in our model, as Pluto's atmosphere is extended with respect to Pluto's radius. The gravitational acceleration at some altitude $z$ is calculated by multiplying the surface $g$ value by the ratio $R^2/(R+z)^2$, where $R$ is Pluto's radius (1187 km; Stern et al. 2015).

For aggregate particles Eq. (8) can also be applied, with $R_f$ in place of $r_p$, a reduced particle density $\rho_p^{agg}$ in place of $\rho_p$ given by

$$\rho_p^{agg} = \rho_p \left(\frac{R_f}{r_m}\right)^{D_f-3} \tag{10}$$

and by taking into account porosity by multiplying Eq. (8) by a factor $\Omega$ (Lavvas et al., 2010)

$$\Omega = \frac{2\beta^2\left(1-\frac{\tanh(\beta)}{\beta}\right)}{2\beta^2+3\left(1-\frac{\tanh(\beta)}{\beta}\right)}\ ,\ \ \beta = \frac{R_f}{\sqrt{\kappa}} \tag{11}$$

where

$$\kappa = \frac{4r_m^2}{18\psi}\frac{3-4.5\psi^{\frac{1}{3}}+4.5\psi^{\frac{5}{3}}-3\psi^2}{3+2\psi^{\frac{5}{3}}}\ ,\ \ \psi = \frac{\rho_p^{agg}}{\rho_p} \tag{12}$$

The standard expression for $K_{i,j}$ (Pruppacher & Klett, 1978; Lavvas et al., 2010) is used in CARMA for both spherical and aggregate particles, where for the latter the radius upon which $K_{i,j}$ depends is $R_f$ instead of $r_p$. In addition, we assume that coagulation is impeded due to particle charging effects, similar to Titan's aerosols, with the coagulation rate reduced by a factor

$$f_c = \frac{\tau}{\exp(\tau)-1},\ \tau = \frac{q^2 r_i r_j}{kT(r_i+r_j)} \tag{13}$$

where $k$ is Boltzmann constant, and $q$ is the charge density (Fuch, 1964). It is unknown what processes could be responsible for charging aerosols in Pluto's atmosphere, though interactions of the haze particles with ambient ions and electrons created from photolysis and interactions with the Solar wind are likely (Cravens and Strobel, 2015), with the low density of the Pluto atmosphere facilitating charge separation due to differential diffusion of free electrons and positive ions.

As an alternative scenario to aggregate particles we also consider spherical haze particles to evaluate the need for low-density aggregates in Pluto's tenuous atmosphere to combat the relatively short sedimentation times. For this simulation we assume that the particles sedimenting into the model domain from above 500 km are spherical but with the same mass as the aggregate particles, such that they have a radius $r_p = 15.87$ nm and 12.6 nm, equivalent to the 10 nm and 5 nm aggregate cases, respectively.

Unlike Titan, where condensation of simple hydrocarbons and nitriles onto aerosols only becomes important in the stratosphere (<110 km), far below the altitudes of haze formation (~1000 km) (Lavvas et al., 2013; Barth, 2015), condensation of such species on Pluto aerosols may occur at altitudes (~200 km, Figure 1) much closer to where the aerosols first form (~500 km, Wong et al., this issue), such that it could impact the observed extinction profiles. However, our current model does not take into account condensation onto fractal aggregates. In addition, as the condensation rates can be comparable to reaction rates governing the chemistry of Pluto's atmosphere (Wong et al., this issue), a coupled photochemistry-microphysics model will be needed to fully explain the observed extinction profiles of chemical species and aerosols. Despite our limitations, however, we will attempt to quantify the effects of condensation in Section 3.2 using a simple analytic model.

We compare our model results to the solar occultation observations of Pluto's atmosphere obtained by the Alice ultraviolet spectrograph onboard *New Horizons* during its flyby (Gladstone et al. 2016). The retrieved line of sight (LOS) optical depth is converted into an extinction coefficient $\alpha$ by using the Abel transform (i.e. Kammer et al. 2013) across the full range of altitudes where data was collected, from ~2000 km above the surface to the surface itself. Uncertainties in $\alpha$ at each altitude are computed using a bootstrap Monte Carlo routine.

Simulated $\alpha$ profiles are derived from the haze particle size and number density distributions calculated by the model by assuming that these particles have the same optical properties as that of tholins, which has been proposed to be the composition of Titan's aerosols, and which possess a complex refractive index of $1.65 + 0.24i$ in the wavelength range (145 – 185 nm) of the data (Khare et al., 1984; Gladstone et al., 2016). $\alpha$ is calculated from the product of the number density of haze particles $N$ and the extinction cross section $\sigma$ of the particles, given by

$$\sigma = Q_e \pi r_p^2 \tag{14}$$

for spheres, and

$$\sigma = Q_e \pi r_m^2 N_m^{2/3} \tag{15}$$

for aggregates, where $Q_e$ is the extinction efficiency calculated using established Mie scattering codes for aggregates (Tomasko et al., 2008) and spheres (Grainger et al., 2004).

## 3. RESULTS AND DISCUSSION

### *3.1. Comparison to Data*

Figure 3 shows the particle number density, expressed in $dN/dLn(R_f)$ for aggregates and $dN/dLn(r_p)$ for spheres, as a function of altitude in Pluto's atmosphere and particle radius for the (clockwise from the top left) 5 nm monomer aggregate, 10 nm monomer aggregate, 10 nm monomer-equivalent spherical, and 5 nm monomer-equivalent spherical haze solutions computed by CARMA. The particle size distribution peaks at ~0.1–0.2 μm below 150 km altitude for the aggregate cases, consistent with particle sizes inferred from forward scattering observations by *New Horizons* (Gladstone et al., 2016), while for the spherical haze cases the particle radii are smaller by about a factor of 4. The low vertical resolution of our model (~25 km at 50 km altitude) is far too coarse to resolve the low altitude (<80 km), few-km-thick, spatially coherent haze layers that were seen in high phase angle observations (Cheng et al., 2015), which are thought to be generated by gravity waves (Gladstone et al., 2016) – a process not included in our model.

Figure 4 compares the extinction coefficients $\alpha$ calculated from our model aggregate (green) and spherical (orange) particle haze results, for both the 10 nm monomer (dash dot line) and 5 nm monomer (dashed line) cases (and the equivalent cases for spherical particles), to that derived from both the ingress (red) and egress (blue) solar occultation observations (Gladstone et al. 2016). We obtain general agreement between the aggregate cases and the data when the downward mass flux of photochemical products is equal to the column-integrated methane destruction rate $\sim 1.2 \times 10^{-14}$ g cm$^{-2}$ s$^{-1}$ computed by the photochemical model of Wong et al. (this issue). This flux is nearly identical to the haze production flux obtained for Titan using similar microphysical models (Lavvas et al., 2010, Larson et al., 2014), despite Titan being exposed to more solar UV photons, which suggests that haze production may be limited by the availability of haze precursors. By comparison, the same downward mass flux applied to the spherical particle cases underestimates $\alpha$ by a factor of ~3.

It should be noted that, given the significance of condensation and surface deposition as sinks for $C_2$ hydrocarbons, which prevents them from participating in further photochemistry in the atmosphere that may convert them to aerosols (Wong et al., this issue), the conversion rate of methane to aerosols is almost certainly less than 100% (Stansberry et al., 1989). Thus, the downward mass flux of haze particles should be less than what is presented here. However, the degree of freedom afforded to aggregates by $r_m$ can alleviate this issue. This can be understood by noting that the $\alpha$ profile for the 5 nm aggregate case is almost identical to the 10 nm monomer case, except that the former is greater than the latter by a factor of ~1.5. An alternate case where $r_m$ is set to 1 nm resulted in $\alpha$ values 5 times greater than that of the 10 nm monomer case (not shown). In other words, decreasing $r_m$ has the effect of increasing $\alpha$, which would compensate for a decrease in $\alpha$ due to a decrease in downward haze particle mass flux. This is because reducing $r_m$ for a fixed $R_f$ increases the porosity of the aggregate and thus decreases its density, which in turn reduces its sedimentation velocity, allowing for more haze particles to stay aloft, increasing $\alpha$. Thus, the uncertainty in conversion efficiency of methane to aerosols prevents us from constraining the aggregate monomer size using occultation observations alone. However, this same degeneracy points strongly to Pluto's haze particles being fractal aggregates, as spherical particles do not have an extra degree of freedom arising from monomers, i.e. decreasing their downward mass flux would just lead to smaller $\alpha$ values.

Figure 5 (top) shows variations in the particle size distribution 1.6 km above the surface of Pluto with changes in the particle charge to radius ratio, given in $e^-/\mu m$. We find that $q = 30$ $e^-/\mu m$ offers the best agreement with the retrieved mean particle size (~0.1-0.2 $\mu m$) from the forward scattering measurements (Gladstone et al., 2016), which is 2-4 times that obtained from observations of Titan aerosols (Lavvas et al., 2010; Larson et al., 2014). This suggests that Pluto's atmosphere may be more amicable to particle charging.

Figure 5 (bottom) shows profiles of extinction coefficients corresponding to the particle charge to radius ratios of Figure 5 (top), and reveals that $\alpha$ is not significantly perturbed when the charge to radius ratio is varied. This results from the similar way the extinction cross section and mass of an aggregate scale with radius, such that a small number of large aggregates would have nearly the same extinction cross section as a large number of small aggregates, provided their total masses were the same. Also, as the density of an aggregate with $D_f$ = 2 is inversely proportional to the aggregate effective radius $R_f$, and $\Omega$ varies slowly with $R_f$ (Lavvas et al., 2010), the sedimentation velocity for aggregates in the kinetics regime is largely independent of $R_f$. Thus, the small differences between the various particle charge to radius ratio cases stem from variations of the extinction efficiency with $R_f$.

Despite the general agreement between our model and data, it is clear that there are deficiencies. For example, the observations show a somewhat steeper $\alpha$ profile compared to what our models produce, characterized by nearly vertical "steps" in between gentler slopes, the latter of which appear to match the slope of our model $\alpha$ profiles (see, for example, the fit between the data and the 5 nm monomer aggregate curve between 150 and 200 km). One possible explanation is that Titan's haze is not a perfect analogy for that of Pluto. On Titan, haze production is effected by both solar UV photons and energetic particles from Saturn's magnetic field, whereas for Pluto this latter energy source either does not exist (Elliot et al., 1989), or is replaced by energetic particles from the solar wind, which could lead to different haze production schemes. In addition, Pluto's atmospheric chemistry may be sufficiently different from Titan such that more nitriles are incorporated into the haze particles by comparison (Wong et al., 2015). Furthermore, the 10 times higher mixing ratio of CO in Pluto's atmosphere (~500 ppm; Lellouch et al., 2011) vis-à-vis Titan (~50 ppm; de Kok et al., 2007) may result in more oxidized haze particles. These effects can lead to different refractive indices for the Pluto haze particles such that using those for tholins may not be appropriate. Quantifying how these subtle differences in haze composition change their refractive indices will require laboratory measurements of analogous materials.

Another possible effect not considered in our model is that of dynamics. Gladstone et al. (2016) showed that vertical velocities of 20 cm $s^{-1}$ at 200 km altitude and <10 cm $s^{-1}$ below 75 km altitude are possible due to gravity waves arising from sublimation and orographic forcing, which are ~10 times larger than the sedimentation velocity above the gravity wave saturation altitude of ~10 km (Figure 2). Though these waves could cause the emergence of layering in the haze, it is unknown how they could affect the overall slope of the extinction profile, as the calculated velocities oscillate between upward and downward motion that may cancel out at longer length scales.

Finally, the fractal dimension of Pluto's hazes may not be exactly 2, considering the range in $D_f$, 1.75 – 2.5, estimated for Titan (Cabane et al., 1993). If $D_f > 2$, then the aggregate case would become more similar to the spherical case. If $D_f < 2$, then the increased porosity of the aggregates would decrease their sedimentation velocities and increase their coagulation rates. As the cross sectional area of the aggregate now grows faster than its mass, this leads to an increase in extinction at lower altitudes, which is not seen in the *New Horizons* observations.

*3.2. Effects of Condensation*

Another possible cause for the discrepancies between model and data is condensation of hydrocarbons and nitriles onto the aggregates, which is necessary to explain the observed extinction due to these chemical species (Wong et al., this issue). As mentioned in Section 2, our current photochemical and microphysical models are decoupled from each other, so a self-consistent treatment of condensation and its impact on both chemical species' mixing ratios and aerosol number density and size distributions is not possible. However, we can still estimate the effects of condensation on $\alpha$ due to changes in particle and monomer size and number density by using the net condensation rates calculated by the photochemical model of Wong et al. (this issue) (Figure 1), scaling relations, and mass balance.

Consider sedimenting particles composed of a condensed layer of hydrocarbons and nitriles around an aggregate aerosol "core". From the net condensation rates, we find a total column-integrated condensation "flux" of ~8.1 × 10$^{-15}$ g cm$^{-2}$ s$^{-1}$, which is about 2/3 of the downward mass flux of haze particles. Thus, it is unlikely for the condensed material to completely inundate the aggregate such that the particle becomes non-porous; instead, it can be assumed that the condensed material merely fills in some of the pores within the aggregate. For simplicity we assume that the condensed mass contributes entirely to increasing the monomer size and/or fusing separate monomers together into single, larger monomers, while keeping $R_f$ and $D_f$ constant. In such a scenario, the mass flux of particles $\Phi_{ij}$ in mass bin $j$ through an altitude level $i$ in the atmosphere would be equal to

$$\Phi_{ij} = \rho_{ij} v_{ij} = \rho_{ij}^{a} v_{ij} + \Phi_{ij}^{c} \tag{16}$$

where $\rho_{ij}$ is the total mass density of particles in bin $j$ at altitude $i$, equal to the mass of a single particle times their number density $N_{ij}$, $\rho_{ij}^{a}$ is the total mass density of just the aggregate core in bin $j$ at altitude $i$, $v_{ij}$ is the sedimentation velocity of the composite particle in bin $j$ at altitude $i$, and $\Phi_{ij}^{c}$ is the net column rate of condensation onto a haze particle in bin $j$ at altitude $i$, calculated by multiplying the net condensation rates in Figure 1 by the thickness of altitude layer $i$. Expanding the mass densities according to Eq. (2) gives

$$N_{ij} \tfrac{4}{3} \pi \rho_p r'_{mij} R_{fij}^2 v_{ij} = N_{ij} \tfrac{4}{3} \pi \rho_p r_{mij} R_{fij}^2 v_{ij} + \Phi_{ij}^{c} \tag{17}$$

where $r'_{mij}$ is the monomer size after the particle passes through altitude $i$ and has changed due to condensation, and $r_{mij}$ is the monomer size before passing through altitude $i$. In other words, we assume that condensation at each altitude level happens instantaneously in the middle of that level, with a condensation rate depending on how fast the particle was travelling through that level. Thus, $r'_{mij} = r_{m(i+1)j}$, where $i$ increases upwards. Isolating the $r_m$'s gives

$$r_{mij} = r_{m(i+1)j} + \frac{3\Phi_{ij}^{c}}{4N_{ij}\pi\rho_p R_{fij}^2 v_{ij}} \tag{18}$$

In Wong et al. (this issue), the condensation rate is assumed to be proportional to the surface area of the haze particles, which were taken to be spheres with radius $R_f$ as a compromise between the increased surface area of aggregates compared to spheres and the limited diffusive pathways a condensate molecule can take to get into the pore spaces of the aggregate. We thus calculate $\Phi_{ij}^c$ by scaling the total net column rate of condensation at altitude $i$ by the total surface area of particles in bin $j$ at that altitude

$$\Phi_{ij}^c = \Phi_i^c \frac{N_{ij} R_{fij}^2}{\sum_j N_{ij} R_{fij}^2} \tag{19}$$

We have thus derived in Eqs. (18)-(19) an iterable expression that can calculate $r_m$ for every altitude and particle bin given $r_m$ at the top of the model and $\Phi_i^c$. Eq. (18) is iterated multiple times per $ij$ pair in order to assert self-consistency, as $v_{ij}$ depends on $r_{mij}$. For a starting $r_m = 10$ nm for all bins at 500 km and $\Phi_i^c$ given by Figure 1, Figure 6 shows the change in the weighted mean of $r_m$ as a function of altitude, weighted by the number density of aggregates, due to condensation by HCN (red), $C_2H_2$ (yellow), $C_2H_4$ (green), $C_2H_6$ (blue), and all of these species combined (black). Minor differences in altitude exist between where each chemical species condense the fastest (i.e. where $r_m$ grows the fastest), but they are all within ~50 km of each other, and all occur above 200 km, which is consistent with Figure 1, while a minor amount of $r_m$ growth also occurs down to 100 km. These two altitudes also coincide roughly with the bottoms of the upper two steep "steps" in the observed extinction profiles (Figure 4), suggesting some connection between condensation and extinction. The third "step" at ~75 km, which is more muted than the top two steps in the egress profile and is absent in the ingress profile correspond to a warmer region of the atmosphere (Figure 1), where the $r_m$ growth curve reverses direction slightly due to net evaporation (not shown in Figure 1). The large increase in $r_m$ occurring near the surface resulting mostly from $C_2H_6$ condensation and rainout is not in the altitude range of the data.

There are three ways in which condensation can affect $\alpha$: (1) By changing the refractive index of the haze particles, (2) by increasing the mass of particles such that their sedimentation velocities change, which in turn affects the equilibrium number density, and (3) by directly changing the cross sections of the aggregates and/or their monomers. The refractive indices of hydrocarbon and nitrile ices in the FUV are not well known (de Bergh et al., 2008; Hendrix et al., 2013) and therefore we cannot speak to the validity of (1), though $\alpha$ is roughly linear with the imaginary part of the refractive index, and so altering its value by a factor of a few could improve the agreement between our model and the data. Options (2) and (3) can be evaluated using scaling relations. Consider again Eq. (16), but only the left and middle terms; expanding the latter out and combining with Eqs. (8), (10)-(12) gives

$$\Phi \propto N r_m^2 \tag{20}$$

where again we have assumed that $R_f$ and $D_f$ stay constant. A similar proportionality can be written for $\alpha$,

$$\alpha \approx \sigma N \propto Q_e r_m^{2/3} N \tag{21}$$

Combining Eqs. (20) and (21) thus gives

$$\alpha \propto Q_e r_m^{-4/3} \Phi \tag{22}$$

Note that, were $R_f$ not held constant, it would enter Eq. (22) as a factor $R_f^{-2/3}$. Using the aggregate scattering code of Tomasko et al. (2008) and fixing $R_f$ to 20 nm and $D_f$ to 2 while changing $r_m$ shows that $Q_e \propto r_m^{0.64}$, so that

$$\alpha \propto r_m^{-0.69} \Phi \tag{23}$$

Including condensation increases $\Phi$ by a factor of 5/3 over the flux corresponding to just sedimenting aerosols, while $r_m$ increases by a factor of 50%, resulting in an increase in $\alpha$ by ~26%. In other words, the increase in mass, and thus extinction efficiency of the particles dominates the decrease in the particle number density caused by an increased sedimentation velocity, resulting in greater extinction overall. This is the opposite of what is seen in the observations, though the effect is small in comparison to the magnitude of the "steps", which decrease $\alpha$ with respect to a smoothly, downward increasing profile by a factor of ~2-3 (see, for example, the difference between the data and the 5 nm monomer aggregate model curve at ~130 km).

The above calculation assumes that the condensed mass only contributes to increasing the aggregate monomer size. Several alternative scenarios exist, which include (1) condensation affecting both $r_m$ and $R_f$, while keeping $N_m$ and $D_f$ fixed, corresponding to the condensate molecules depositing on all available aggregate surfaces equally, and (2) condensation affecting both $R_f$ and $N_m$, while keeping $r_m$ and $D_f$ fixed, corresponding to the condensate forming a thin shell around the aggregate, leaving pore spaces unaffected. However, after repeating our above procedure (Eqs. 16-23) for these alternate scenarios, we find that none of them decrease $\alpha$ after condensation occurs. The scenarios where $D_f$ changes were not investigated, as the Tomasko et al. (2008) model has not been tested for $D_f \neq 2$. Nonetheless, the case where condensation affects both $r_m$ and $D_f$ and not $R_f$ and $N_m$, corresponding to the infilling of the pore spaces of the aggregate offers another likely possibility, but will require a more rigorous treatment of aggregate extinction.

The simplicity of Eqs. (16)-(23) begets several caveats. For example, coagulation of the composite particles is not taken into account, the rate of which would change due to the different mass, size, and number density of the particles. The condensation rates themselves could also change due to variations in aggregate surface area, sticking coefficient, and number density. In addition, we have assumed that the condensation of $C_2$ hydrocarbons and HCN is a separate process from haze production, whereas in reality both the haze and the condensates arise from methane photolysis. Thus, if the condensation flux were fixed (~$8.1 \times 10^{-15}$ g $cm^{-2}$ $s^{-1}$; necessary to reproduce the observations of chemical abundances; Wong et al. this issue), then the actual haze production rate would need to be lower by a factor of 3 such that their combined rate equals the methane photolysis rate. As per our earlier discussion regarding the degeneracy between monomer radius and haze production rate in determining $\alpha$, such a decrease in haze production can be remedied by decreasing the monomer radius. Therefore, a possible scenario is that the haze aggregates are produced with a rate equal to 1/3 of the methane photolysis rate in mass, and with monomers ~1 nm in radius; the aggregates then act as condensation sites for $C_2$ hydrocarbons and HCN as they descend through the atmosphere, eventually accreting enough material such that their monomers grow to ~10 nm near the surface. Modeling of this process will require a self-consistent aggregate microphysical model that takes into account coagulation, condensation, and their feedbacks on photochemistry.

We have shown that the hazes of Pluto are likely composed of fractal aggregates similar to those of Titan, with their distribution shaped by sedimentation and coagulation. The mean haze particle sizes and extinction profiles calculated by our microphysical model are in general agreement with forward scattering and solar occultation observations obtained by *New Horizons*. Discrepancies between the model results and the observations could be due to subtle differences in composition between the Pluto hazes and the tholins derived in Earth laboratories caused by different chemistry, formation processes, and nucleation and condensation of HCN and $C_2$ hydrocarbons onto the haze particles.

**Acknowledgements**

This research was supported in part by a grant from the New Horizons Mission. YLY and RLS were supported in part by the Cassini UVIS program via NASA Grant JPL.1459109, NASA NNX09AB72G grant to the California Institute of Technology. PG was supported in part by an RTD grant from JPL.

## REFERENCES

Bardeen, C. G., Toon, O. B., Jensen, E. J., et al., 2008. Numerical simulations of the three dimensional distribution of meteoric dust in the mesosphere and upper stratosphere. J. Geophyis. Res. 113 (D17202), 15pp.

Barth, E. L., 2014. Haze particles and condensation in Pluto's atmosphere explored through microphysical modeling. American Geophysical Union, Fall Meeting 2014, P33B-4036.

Barth, E. L., 2015. Condensation of trace species to form ice layers in Titan's stratosphere. Division of Planetary Sciences Meeting 47, 300.02.

Cabane, M., Rannou, P., Chassefiere, E., et al., 1993. Fractal aggregates in Titan's atmosphere. Planet. Space Sci. 41, 257-267.

Cheng, A. F., Gladstone, G. R., Summers, M., et al. 2015. Discovery of hazes in Pluto's atmosphere. American Astronomical Society, Division of Planetary Sciences Meeting 47, 105.02.

Cravens, T. E., Strobel, D. F., 2015. Pluto's solar wind interaction: Collisional effects. Icarus 246, 303-309.

Cruikshank, D. P., Brown, R. H., 1986. Satellites of Uranus and Neptune, and the Pluto-Charon system. In: Satellites. Eds. Burns, J. A., Matthews, M. S. University of Arizona Press, Tucson, AZ, USA, pp. 836-873.

Dias-Oliveira, A., Sicardy, B., Lellouch, E., et al., 2015. Pluto's atmosphere from stellar occultations in 2012 and 2013. Astrophys. J. 811, 53.

de Bergh, C., Schmitt, B., Moroz, L. V., et al., 2008. Laboratory data on ices, refractory carbonaceous materials, and minerals relevant to transneptunian objects and centaurs. In: The Solar System beyond Neptune. Eds. Barucci, M. A., Boehnhardt, H., Cruikshank, D. P. et al. University of Arizona Press, Tucson, AZ, USA, pp. 483-506.

de Kok, R., Irwin, P. G. J., Teanby, N. A., et al., 2007. Oxygen compounds in Titan's stratosphere as observed by Cassini CIRS. Icarus 186, 354-363.

Elliot, J. L., Dunham, E. W., Bosh, A. S., et al., 1989. Pluto's atmosphere. Icarus 77, 148-170.

Elliot, J. L., Ates, A., Babcock, B. A., et al., 2003. The recent expansion of Pluto's atmosphere. Nature 424, 165-168.

Fuch, N. A., 1964. The Mechanics of Aerosols. Pergamon Press, New York (translated by R.E. Daisley and M. Fuchs).

Gladstone, G. R., Stern, S. A., Ennico, K., et al., 2016. The atmosphere of Pluto as observed by New Horizons. Science 351, 1280.

Grainger, R. G., Lucas, J., Thomas, G. E., et al., 2004. Calculation of Mie derivatives. Appl. Opt. 43, 5386-5393.

Gulbis, A. A. S., Emery, J. P., Person, M. J., et al., 2015. Observations of a successive stellar occultation by Charon and graze by Pluto in 2011: Multiwavelength SpeX and MORIS data from the IRTF. Icarus 246, 226-236.

Hendrix, A. R., Domingue, D. L., Noll, K. S., 2013. Ultraviolet properties of planetary ices. In: The science of Solar System ices. Eds. Gudipati, M. S., Castillo-Rogez, J. Springer-Verlag, New York, NY, USA.

Jacobson, M. Z., Turco, R. P., 1994. Modeling coagulation among particles of different composition and size. Atmos. Environ. 28, 1327-1338.

Kammer, J. A., Shemansky, D. E., Zhang, X., Yung, Y. L., 2013. Composition of Titan's upper atmosphere from Cassini UVIS EUV stellar occultations. Planet. Space Sci. 88, 86-92.

Khare, B. N., Sagan, C., Arakawa, E. T., et al., 1984. Optical constants of organic tholins produced in a simulated Titanian atmosphere: From soft X-ray to microwave frequencies. Icarus 60, 127-137.

Larson, E. J. L., Toon, O. B., Friedson, A. J., 2014. Simulating Titan's aerosols in a three dimensional general circulation model. Icarus 243, 400-419.

Larson, E. J. L., Toon, O. B., West, R. A., Friedson, A. J., 2015. Microphysical modeling of Titan's detached haze layer in a 3D GCM. Icarus 254, 122-134.

Lavvas, P., Yelle, R. V., Griffith, C. A., 2010. Titan's vertical aerosol structure at the Huygens landing site: Constraints on particle size, density, charge, and refractive index. Icarus 210, 832-842.

Lavvas, P., Yelle, R. V., Koskinen, T., et al., 2013. Aerosol growth in Titan's ionosphere. Proc. Natl. Acad. Sci. U.S.A. 110, 2729-2734.

Lellouch, E., de Bergh, C., Sicardy, B., et al., 2011. High resolution spectroscopy of Pluto's atmosphere: Detection of the 2.3 μm $CH_4$ bands and evidence for carbon monoxide. Astronom. and Astrophys. 530, L4.

Olkin, C. B., Young, L. A., French, R. G., et al., 2014. Pluto's atmospheric structure from the July 2007 stellar occultation. Icarus 239, 15-22.

Owen, T. C., Roush, T. L., Cruikshank, D. P., et al., 1993. Surface ices and atmospheric composition of Pluto. Science 261, 745-748.

Person, M. J., Dunham, E. W., Bosh, A. S., et al., 2013. The 2011 June 23 stellar occultation by Pluto: Airborne and ground observations. Astron. J. 146, 83.

Pruppacher, H. R., Klett, J. D., 1978. Microphysics of Clouds and Precipitation. Reidel Publishing Caompany, Dordrecht.

Rannou, P., Durry, G., 2009. Extinction layer detected by the 2003 star occultation on Pluto. J. Geophys. Res. 114, E11013, 8pp.

Stansberry, J. A., Lunine, J. I., Tomasko, M. G., 1989. Upper limits on possible photochemical hazes on Pluto. Geophys. Res. Lett. 16, 1221-1224.

Stansberry, J. A., Lunine, J. I., Hubbard, W. B., et al., 1994. Mirages and the nature of Pluto's atmosphere. Icarus 111, 503-513.

Stern, S. A., Bagenal, F., Ennico, K., et al., 2015. The Pluto system: Initial results from its exploration by New Horizons. Science 350, 292.

Tomasko, M. G., Doose, L., Engel, S., et al., 2008. A model of Titan's aerosols based on measurements made inside the atmosphere. Planet. Space Sci. 56, 669-707.

Toon, O. B., Turco, R. P., Westphal, D., et al., 1988. A multidimensional model for aerosols: Description of computational analogs. J. Atmos. Sci. 45, 2123-2143.

Toon, O. B., McKay, C. P., Griffith, C. A., Turco, R. P., 1992. A physical model of Titan's aerosols. Icarus 95, 24-53.

Turco, R. P., Hamill, P., Toon, O. B., et al., 1979. A one-dimensional model describing aerosol formation and evolution in the stratosphere: I. Physical processes and mathematical analogs. J. Atmos. Sci. 36, 699-717.

West, R. A., Smith, P. H., 1991. Evidence for aggregate particles in the atmospheres of Titan and Jupiter. Icarus 90, 330-333.

White, F. M., 1991. Viscous Fluid Flow. McGraw-Hill, New York, USA.

Wolf, E. T., Toon, O. B., 2010. Fractal organic hazes provided an ultraviolet shield for early Earth. Science 328, 1266-1268.

Wong, M. L., Yung, Y. L., Gladstone, G. R., 2015. Pluto's implications for a snowball Titan. Icarus 246, 192-196.

Wong, M. L., Fan, S., Gao, P., et al., 2016. Photochemical model of hydrocarbons on Pluto and comparison to New Horizons observations. Icarus. Accepted. DOI: 10.1016/j.icarus.2016.09.028

Young, E. F., French, R. G., Young, L. A., et al., 2008. Vertical structure in Pluto's atmosphere from the 2006 June 12 stellar occultation. Astron. J. 136, 1757-1769.

Zhu, X., Strobel, D. F., Erwin, J. T., 2014. The density and thermal structure of Pluto's atmosphere and associated escape processes and rates. Icarus 228, 301-314.

## FIGURES

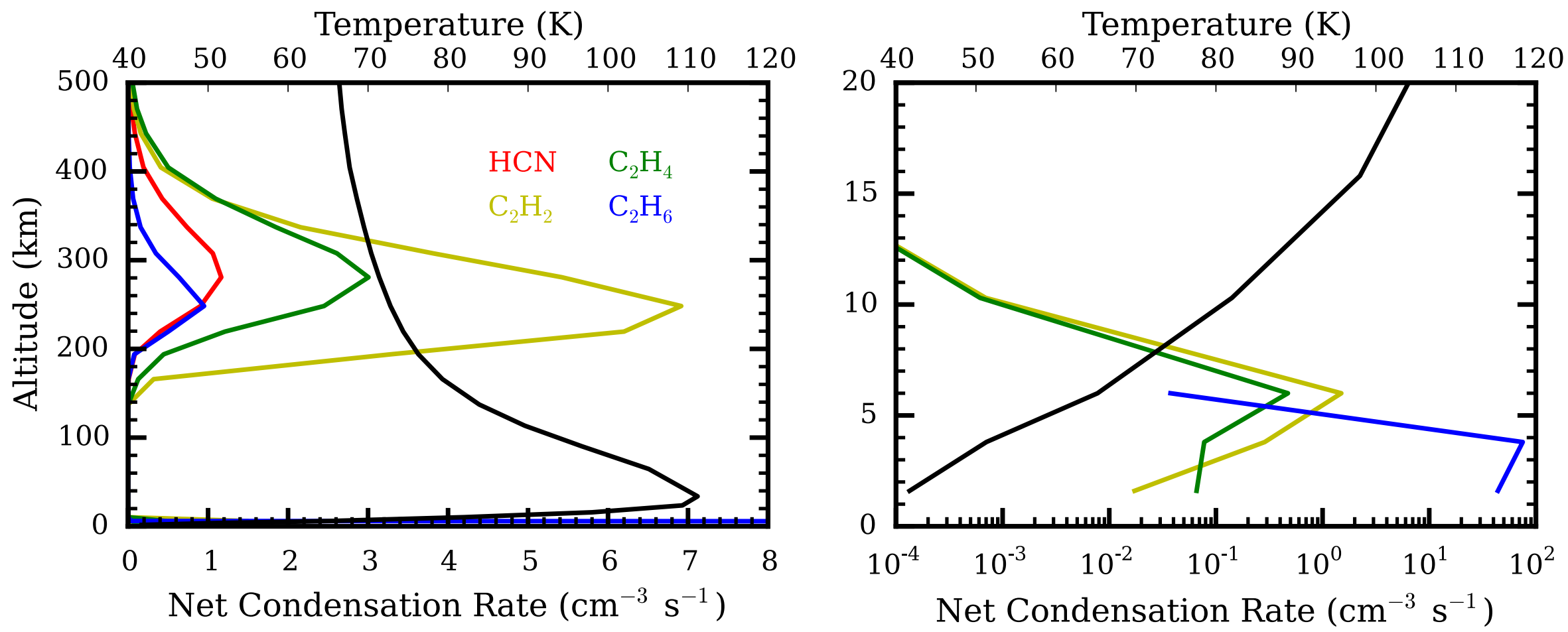

**Figure 1.** (Left) The temperature (black) and net condensation rate profiles for HCN (red), $C_2H_2$ (yellow), $C_2H_4$ (green), and $C_2H_6$ (blue) taken from the photochemical model results of Wong et al. (this issue). (Right) The same as the left plot but zoomed into the lower 20 km of the atmosphere, where $C_2H_6$ condensation dominates over the other species. Note the different bottom x-axis scale between the two plots.

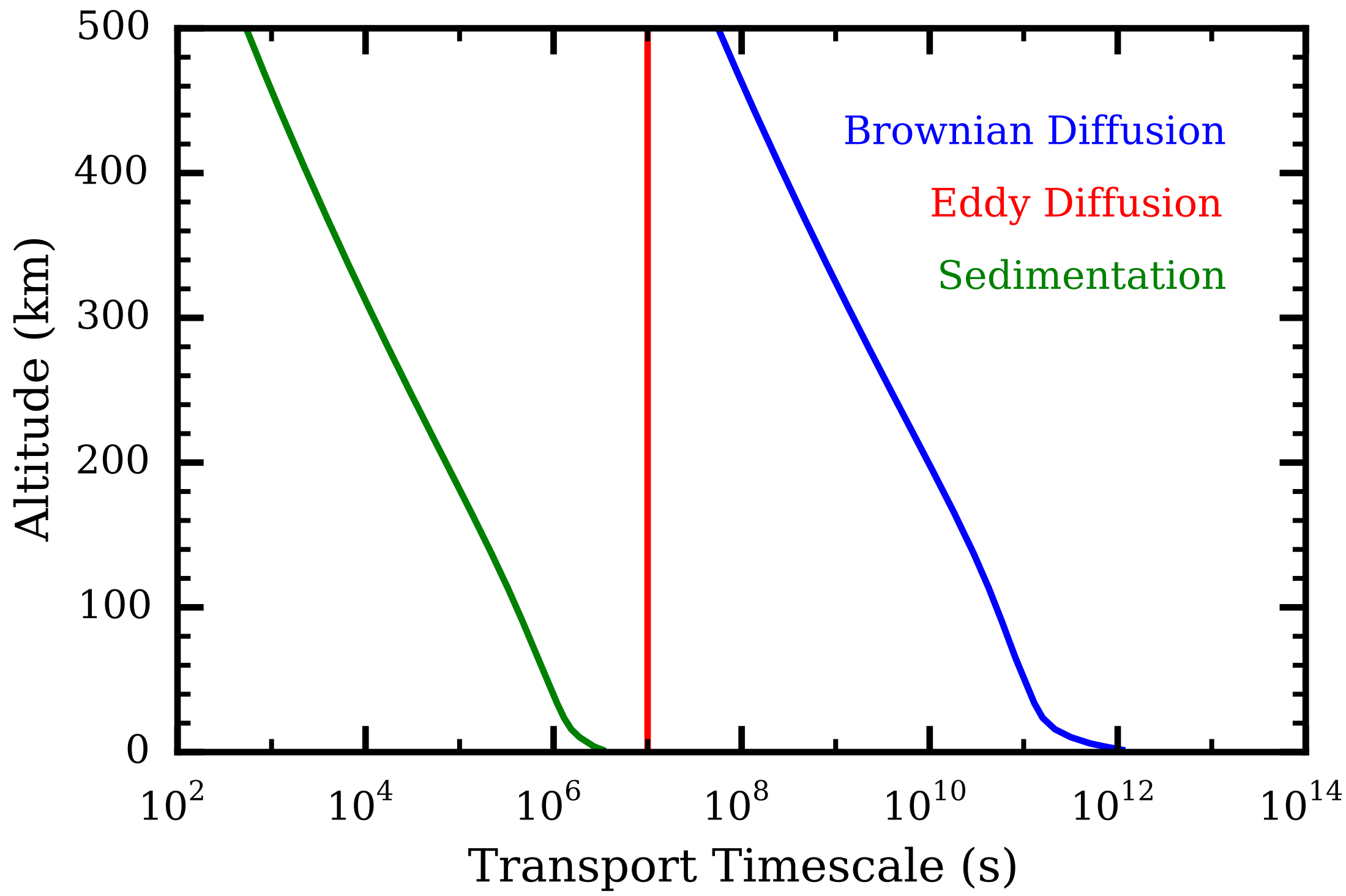

**Figure 2.** The time needed to traverse 1 km in the Pluto atmosphere as a function of altitude for an aggregate particle with $R_f = 0.1$ $\mu$m, $r_m = 10$ nm, and $D_f = 2$ undergoing (blue) Brownian diffusion, (red) eddy diffusion, or (green) sedimentation.

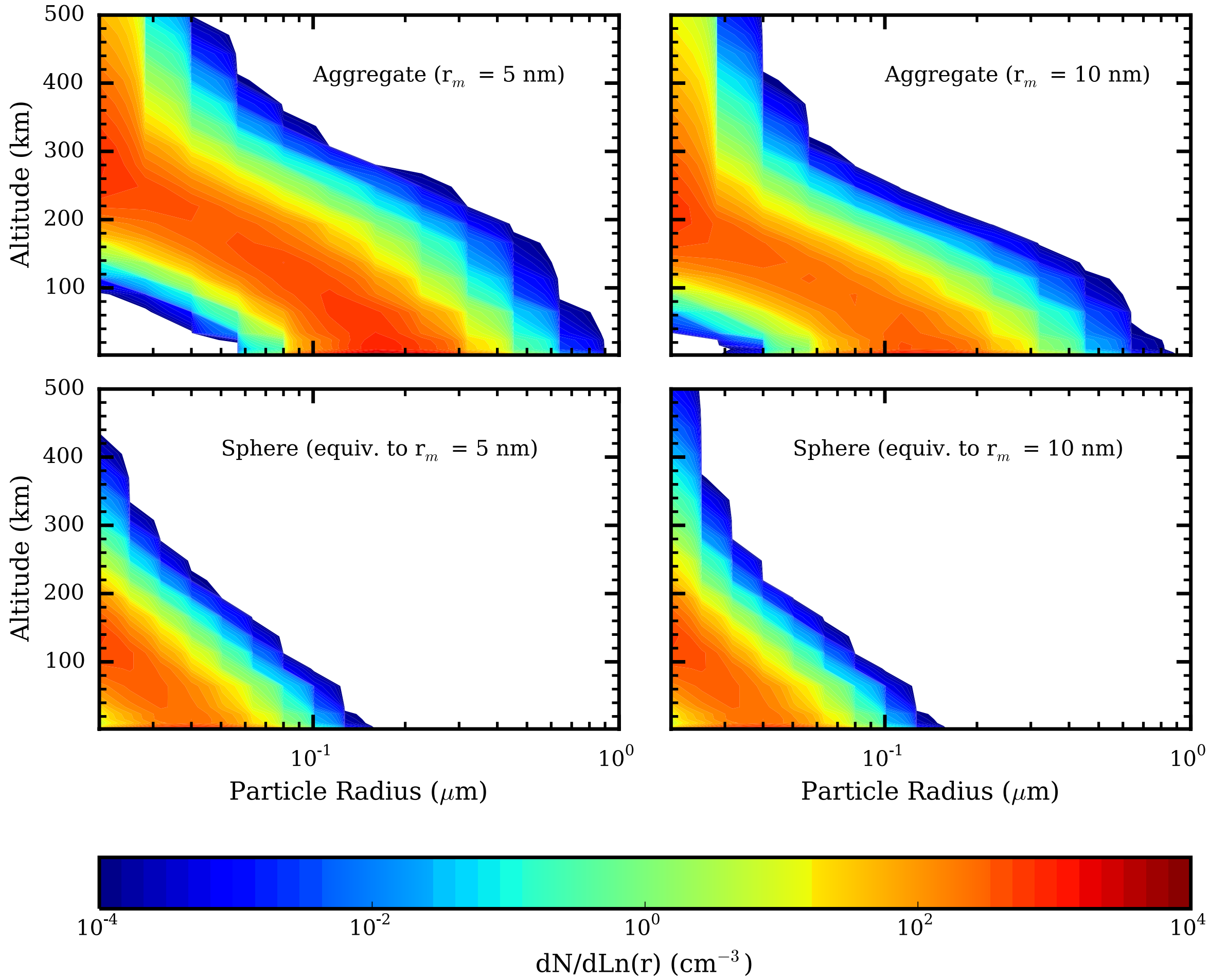

**Figure 3.** Particle number density as functions of altitude and particle radius for the (clockwise from the top left) 5 nm monomer aggregate, 10 nm monomer aggregate, 10 nm monomer-equivalent spherical, and 5 nm monomer-equivalent spherical haze solutions computed by CARMA. Particle radius refers to the true radius $r_p$ for spheres and effective radius $R_f$ for aggregates. Number density is expressed in $dN/dLn(r_p)$ for spheres and $dN/dLn(R_f)$ for aggregates.

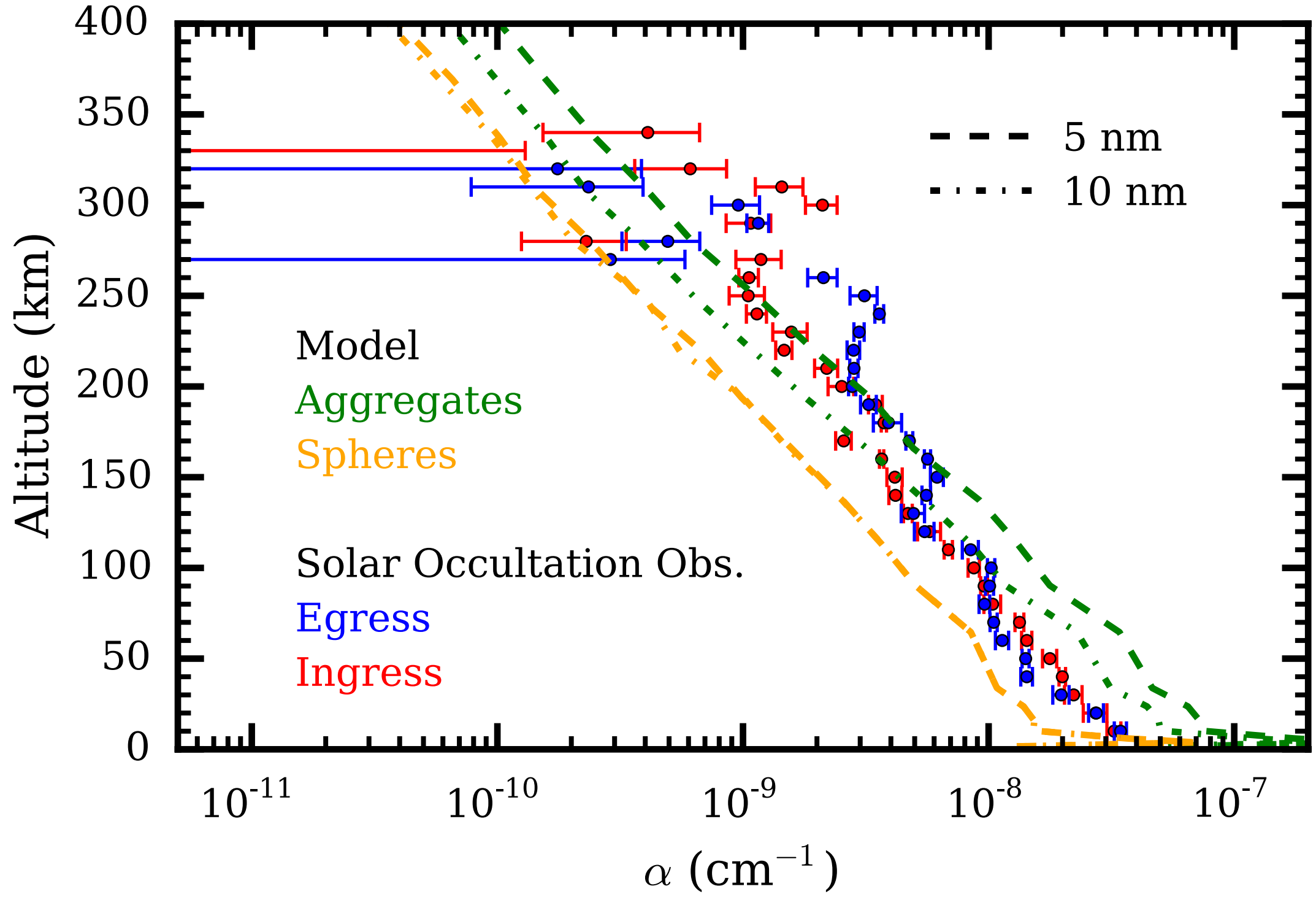

**Figure 4.** The extinction coefficients $\alpha$ as a function of altitude calculated from our model aggregate (green) and spherical (orange) particle haze results, for both the 10 nm monomer (dash dot line) and 5 nm monomer (dashed line) cases (and the equivalent cases for spherical particles), compared to that derived from the ingress (red) and egress (blue) solar occultation observations of Gladstone et al. (2016).

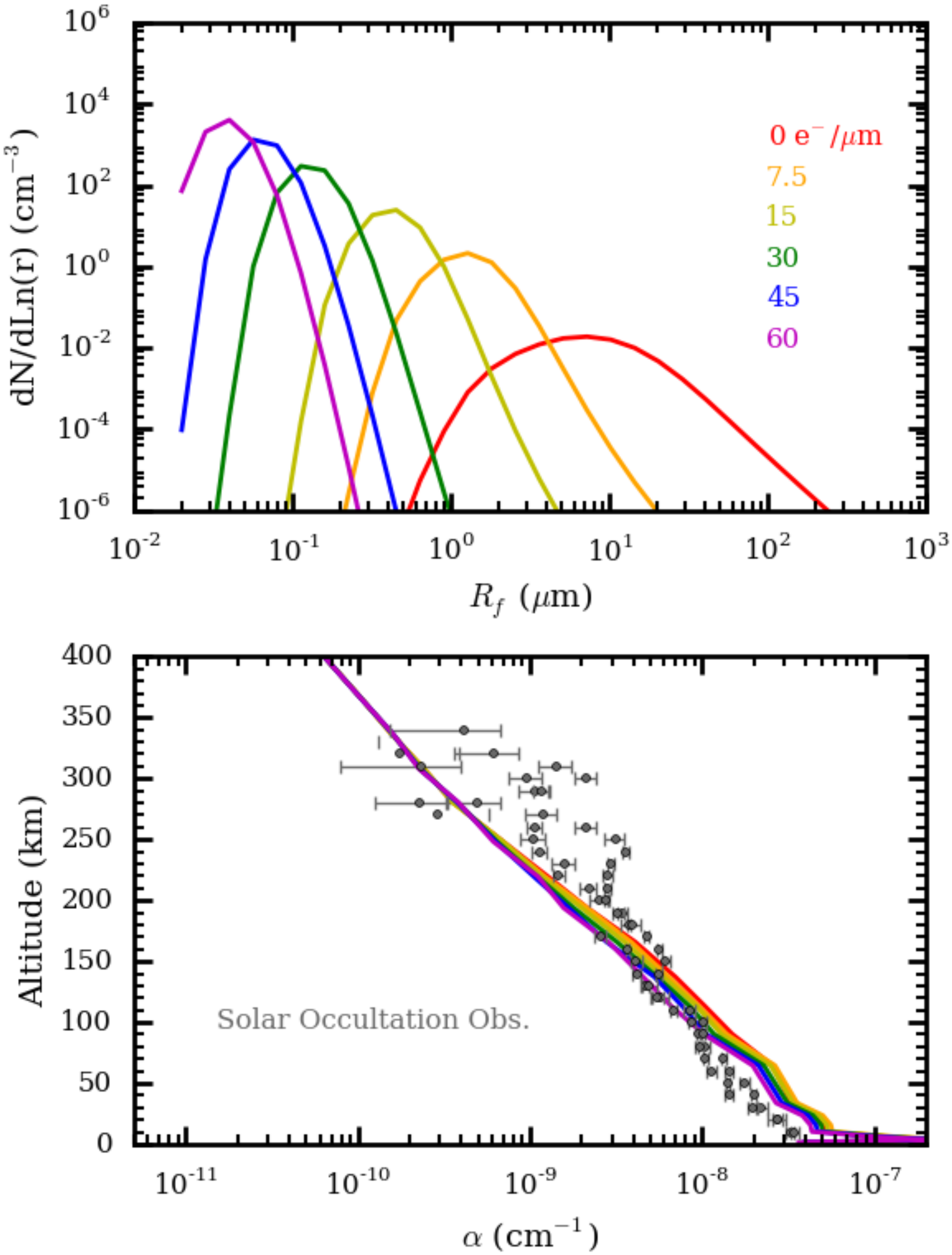

**Figure 5.** (Top) Particle size distributions for charge to particle radius ratios of 0 (red), 7.5 (orange), 15 (yellow), 30 (green), 45 (blue), and 60 $e^{-}/\mu m$ (magenta). (Bottom) Extinction coefficients $\alpha$ corresponding to the different charge to particle radius ratios compared to the ingress and egress *New Horizons* solar occultation observations (gray points; ingress and egress data are not distinguished from each other).

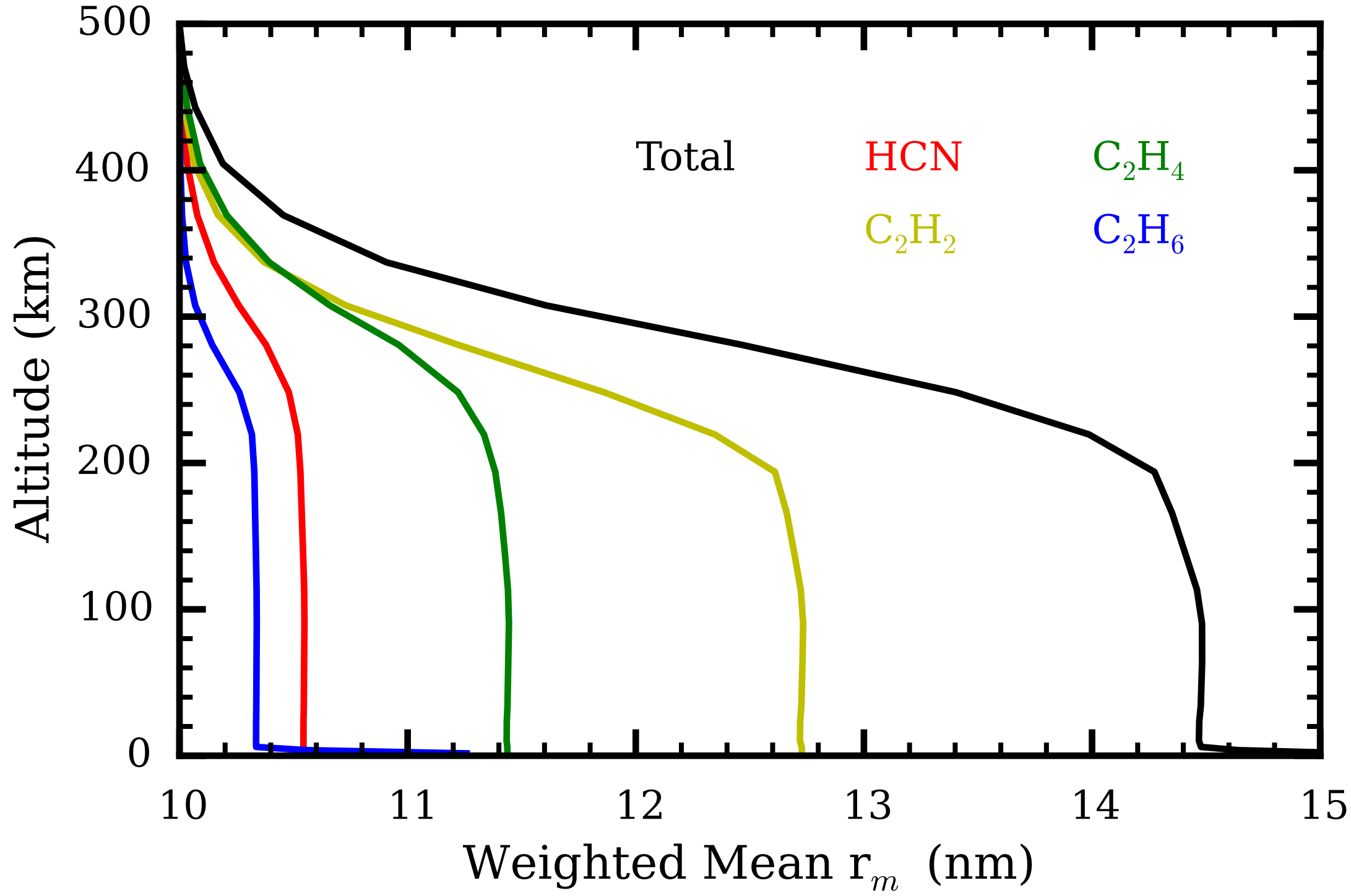

**Figure 6.** The change in mean monomer radius $r_m$ for the 10 nm monomer aggregate case, weighted by the particle number density, as a function of altitude due to condensation of HCN (red), $C_2H_2$ (yellow), $C_2H_4$ (green), and $C_2H_6$ (blue). The total change in weighted mean $r_m$ is shown in black.